\documentstyle[preprint,aps]{revtex}
\newcommand{\sgn}{\mbox{sgn}}
\newcommand{\nonun}{\nonumber}

\newcommand{\bnor}{\left(\frac{\pi}{L}\right)^2}
\newcommand{\bnr}{\frac{\pi}{L}}

\begin{document}


\draft

\title{%
Chiral Tomonaga-Luttinger Liquids and \\
the Calogero-Sutherland Model with
Boundaries
\footnote[2]{%
to appear in the Proceedings of the
Pacific Conference on Condensed Matter Theory,
``Complex Materials and Strongly Correlated Systems'',
Seoul, 2 - 5 December 1995,
which will be published as Supplement to the Journal of the
Korean Physical Society.
}\\
}
\author{%
Takashi Yamamoto
}
\address{%
Yukawa Institute for Theoretical Physics,
Kyoto University  \\
Kyoto 606, Japan}
\author{%
Norio Kawakami
}
\address{%
Department of Applied Physics,
and
Department of Material and Life Science,
Osaka University  \\
Suita, Osaka 565, Japan}
\author{%
Sung-Kil Yang
}
\address{%
Institute of Physics,
University of Tsukuba  \\
Ibaraki 305, Japan}

\maketitle

\begin{abstract}
Applying boundary conformal field theory we study
the low-energy critical behavior of the Calogero-Sutherland model
of $BC_N$-type.  The universality class of the model
is found to be a {\it chiral} Tomonaga-Luttinger liquid.
Various correlation exponents depending on the interaction strength
are obtained.
\end{abstract}

\newpage

\section{Introduction}

\noindent
Recently the one-dimensional quantum integrable models
with long-range interactions have received renewed interest.
The original model of this type
is the Calogero-Sutherland (CS) model \cite{Calogero,CS}
which consists of $N$ particles repelling each other
via inverse-square interactions on a circle of circumference $L$.
The Hamiltonian is given by
\begin{eqnarray}
{\cal H}_{A}
=
-\sum_{j=1}^N
\frac{\partial^2}{\partial q_j^2}
&+&
2\lambda(\lambda-1)
\bnor
\sum_{1\leq j<k\leq N}
\frac{1}{\displaystyle{
\sin^2\bnr(q_j-q_k)}},
 \label{a-tri-hamiltonian}
\end{eqnarray}
where $(q_1,q_2,\cdots,q_N)$
denote particle coordinates
and $\lambda$ is a coupling constant.
The CS model has various significant features:
(i) This model has relatively simple eigen functions
which enable us to perform the exact computation of physical
quantities such as dynamical correlation functions \cite{ha}.
(ii) The energy spectrum is
shown to be reproduced exactly with
the use of the
asymptotic Bethe-ansatz (ABA) method \cite{CS}.
(iii) The excitations are described by quasiparticles
that obey exclusion statistics \cite{ex}.
(iv) The CS model is a typical example
of the Tomonaga-Luttinger liquid \cite{K-Y91}.
These simple, but non-trivial  natures of the CS model
originate from the integrability of the system and
affirm its paradigmatic role
as the anyonic analog of the free boson or fermion gas.
In addition the CS model is related to various branches of physics
and contains many interesting aspects in mathematical physics
(see for reviews \cite{rev-CS} and references therein).
Especially it is established that
this model is the universal Hamiltonian for disordered
systems (typifying quantum chaos)\cite{alt}.

In the original formulation of the CS model,
the interaction among the particles is simply given in a  pairwise form
depending on the difference of particle coordinates.
This simple pairwise interaction is understood as reflecting the underlying
structure of the root system of type $A_{N-1}$. Thus the integrability of the
model has its connection with the theory of Lie algebras. Similarly to the case
of integrable soliton theories which are systematically constructed on the
basis of the Lie algebra, the integrability still holds for
many-body systems with inverse-square potentials which are
obtained by means of
root systems associated to the general Weyl groups \cite{O-Pa}.
Although these models have nice properties which reflect
the Weyl group symmetries, it seems that little attention from the physics
point of view has been paid to the models except the $A_{N-1}$-CS model.

More recently, it has been revealed that
the so-called CS model of $BC_N$-type
($BC_N$-CS model)
which is the most general one among the extended CS models
is relevant to one-dimensional physics with boundaries,
for example,
the boundary quantum sine-Gordon model \cite{K-S94}
and the Dorokhov-Mell-Pereyra-Kumar equation \cite{B-R} which describes
evolution of an ensemble of quasi one-dimensional
disordered wires.
Therefore it is now important to clarify the critical properties of
the $BC_N$-CS model toward our further understanding
in one-dimensional physics including boundary effects.

Our purpose in this contribution is to analyze
the long-distance critical properties
of the $BC_N$-CS model\cite{y-k-y},
following the earlier consideration of the $A_{N-1}$-CS model \cite{K-Y91}.
In particular it will be shown that
the critical behavior of the $BC_N$-CS model is described by
$c=1$ Gaussian conformal field theory (CFT)
with {\it boundaries} \cite{Cardy}
in which we have only left (or right) moving sector of CFT.
Hence the universality class of the $BC_N$-CS model
will be identified as
a {\it chiral} Tomonaga-Luttinger liquid \cite{ch-TLL}.

\section{$BC_N$-CS model
and asymptotic Bethe-ansatz spectrum}

\noindent
We recall the $BC_N$-CS model \cite{O-Pa}.
The $BC_N$-CS model
is related to the root system of type $BC_N$
and invariant under the action of the Weyl group of type $B_N$.
That is, the model is invariant under
the exchange of particle coordinates
and the sign change of coordinates.
The latter is regarded as the exchange of the particle
and its mirror image particle with respect to the origin.
The model has the periodic Hamiltonian
\begin{eqnarray}
{\cal H}_{BC}
=
-\sum_{j=1}^N
\frac{\partial^2}{\partial q_j^2}
&+&
2\lambda(\lambda-1)
\bnor
\sum_{1\leq j<k\leq N}
\left\{
\frac{1}{\displaystyle{
\sin^2\bnr(q_j-q_k)}}
+\frac{1}{\displaystyle{
\sin^2\bnr(q_j+q_k)}}
\right\}
\nonun
\\
&+&
\mu(\mu-1)
\bnor
\sum_{j=1}^N \frac{1}{\displaystyle{
\sin^2\bnr q_j}}
+
\nu(\nu-1)
\bnor
\sum_{j=1}^N \frac{1}{\displaystyle{
\cos^2\bnr q_j}},
 \label{b-tri-hamiltonian2}
\end{eqnarray}
where
$\lambda, \mu, \nu$ $(> 0)$
are coupling constants.
It is clearly seen that the Hamiltonian
(\ref{b-tri-hamiltonian2})
is invariant
under the action of the Weyl group of type $B_N$,
however, is not translationally invariant.
Therefore
the total momentum
is not a good quantum number for the $BC_N$-CS model.

The interaction terms
which violate translationally invariance
are required by
invariance
under the
action of the Weyl group
of type $B_N$.
These terms have clear physical interpretation:
the term
$1/\sin^2(\pi/L)(q_j+q_k)$
represents the two-body interaction
between the $j$-th particle
and
the mirror-image of the $k$-th particle ($j\ne k$).
The terms
$1/\sin^2(\pi/L)q_j$ and $1/\cos^2(\pi/L)q_j$
can be regarded as
the potential
due to {\it impurities}
located at the origin.

The spectrum of the Hamiltonian (\ref{b-tri-hamiltonian2}) of the
$BC_N$-CS model has already been obtained \cite{Nao94}.
Let us present the result in the ABA form.
The energy eigenvalue of the system then
takes the non-interacting form
\begin{equation}
\label{one-exc-ene}
E_N
=
\sum_{j=1}^N
{k_j}^2,
\end{equation}
where pseudomomenta $k_j$'s
satisfy
$k_1>k_2>\cdots>k_N>0$ and obey
the ABA equations
\begin{eqnarray}
\label{exci-rapidi}
k_j L
&=&
2\pi I_j
+
\pi
(\lambda-1)
\sum_{l=1,l\ne j}^N
\left\{
\sgn(k_j-k_l)
+
\sgn(k_j+k_l)
\right\}
\nonumber
\\
& &
+
\pi
(\mu-1)
\sgn(k_j)
+
\pi
(\nu-1)
\sgn(k_j),\hskip10mm j=1,\cdots,N,
\end{eqnarray}
with $\sgn(x)=1$ for $x>0$, $=0$ for $x=0$ and $=-1$ for $x<0$.
The quantum numbers
$I_j\in \mbox{{\bf Z}}_{>0}\ (j=1,\cdots,N)$
characterize the excited states. Notice that the form of Bethe-ansatz
equations (\ref{exci-rapidi}) is common to the Bethe-ansatz solvable
models with boundaries, for example, the nonlinear Schr\"{o}dinger equation
on the half line\cite{gaudin71}
and the $XXZ$ model with
open boundary conditions \cite{H-Q-B87}.

It is immediate to solve the ABA equations (\ref{exci-rapidi}),
obtaining
\begin{eqnarray}
k_j
&=&
\frac{2\pi}{L}
\left[
I_j
-
\left(
N-j+1
\right)
\right]
+
k_j^{(0)}, \hskip10mm j=1,\cdots,N,
\end{eqnarray}
where
\begin{equation}
\label{gra-rapidi}
k_j^{(0)}
=
\frac{2\pi}{L}
\left[
\lambda(N-j)+\frac{\mu+\nu}{2}
\right].
\end{equation}
We see that the ground state is specified by the quantum numbers
$I_j^{(0)}=N-j+1,\ (j=1,\cdots,N)$.
Thus, we get the Fermi point $I_1^{(0)}=N$
and the Fermi momentum
$k_F=\max\{k_j^{(0)}\}=2\pi\lambda N/L+\pi(\mu+\nu-2\lambda)/L$.
It is important to note that the Fermi surface of the
$BC_N$-CS model consists of a single point.

For later use, we shall evaluate the Fermi velocity of the
elementary excitations for the $BC_N$-CS model.
The Fermi velocity of the $BC_N$-CS model cannot be
determined from the dispersion relation,
because the momentum is not a good quantum number.
In order to circumvent this apparent difficulty
we assume that CFT for the $BC_N$-CS model has the central charge
$c=1$. This assumption seems to be legitimate since
the $A_{N-1}$-CS model which may be regarded as the bulk
counterpart of the $BC_N$-CS model
is described in terms of $c=1$ CFT \cite{K-Y91}.
Having the value of the central charge we can determine the Fermi velocity
$v_{\mbox{{\tiny F}}}$ from the low-temperature
expansion  formula of the free energy $F(T)$,
\begin{equation}
\label{ddd}
 F(T)
\simeq
 F(T=0)
-
\frac{\pi T^2}{6v_{\mbox{{\tiny F}}}}c,
\end{equation}
where
$T$ is the temperature. Now, applying the method of Yang and Yang
for the elementary excitation at finite temperatures \cite{themodynamics}
to the ABA formula (\ref{exci-rapidi}),
one can perform the low-temperature expansion
of the free energy $F(T)$ of the $BC_N$-CS model. We get
\begin{equation}
\label{lte}
 F(T)
\simeq
 F(T=0)
-
\frac{\pi T^2}{6(4\pi \lambda d)},
\end{equation}
where $d=N/L$ is the particle density.
Then comparing (\ref{lte}) and (\ref{ddd}) with $c=1$
we obtain $v_{\mbox{{\tiny F}}}=4\pi\lambda d$.
In what follows the validity of our assumption will be confirmed.

\section{Finite-size scaling analysis}

\noindent
When the principle of conformal invariance is applicable we have a
powerful way of obtaining the exponents of correlation
functions in the long-wavelength limit.
As is mentioned in the previous section
the Fermi surface of the
$BC_N$-CS model consists of a single point.
This implies that the
low-energy critical behavior
of the $BC_N$-CS model
will be effectively
described by
a left (or right)-moving sector of
CFT. Hence we expect that boundary CFT
will play a role in our study of the $BC_N$-CS model.
To begin with,
we summarize several fundamental formulas in boundary CFT
\cite{Cardy} which will be needed to analyze the energy spectrum.

Using conformal invariance under {\it free boundary conditions} \cite{B-C-N}
we have the finite-size
scaling form of the ground-state energy
\begin{equation}
E^{(0)}
=
L\epsilon^{(0)} + 2f - \frac{\pi v_{\mbox{{\tiny F}}}}{24L}c \ ,
\label{fsscft}
\end{equation}
where $\epsilon^{(0)}$ and $f$ are, respectively,
the bulk limits of the ground-state energy density
and the boundary energy.
The central charge $c$
which labels the universality class of the system
appears as the universal amplitude of the $1/L$ term in (\ref{fsscft}).

Consider a critical system on the half-plane
$\{(y,\tau)\in
\mbox{{\bf R}}_{\geq 0}\times \mbox{{\bf R}}\}$
with a surface at $y=0$.
($\tau$ means the imaginary time.)
Let ${\cal A}(y,\tau)$ be a local operator.
We consider its two-point correlation function
$G(y_1,y_2,\tau)=
\langle{\cal A}(y_1,\tau_1){\cal A}(y_2,\tau_2)\rangle$,
which is a function of $\tau=\tau_1-\tau_2$ because
of translational invariance along the surface. For
$|\tau|\gg y_1,\, y_2$, the two-point function $G$ has
the asymptotic form
\begin{equation}
\label{long-time-asymp}
G(y_1,y_2,\tau)
\sim
\frac{1}{\tau^{2x_b}},
\end{equation}
where $x_b$ is called the boundary critical exponent.

The boundary critical exponents $x_b$ can be read off from
the scaling behavior of the excitation energy \cite{Cardy}
\begin{equation}
E-E^{(0)}
= \frac{\pi v_{\mbox{{\tiny F}}}}{L}x_b ,
\label{ex-corr}
\end{equation}
where $E$ is the excitation energy.
The value of $x_b$ is generically distinct from that of
the bulk exponent for a corresponding scaling operator. Note also that,
in terms of CFT, the bulk exponent is expressed as
the sum of left and right conformal weights,
while the boundary exponent is equal to the left (or right)
conformal weight.

Let us now calculate the finite-size corrections to the spectrum of
the $BC_N$-CS model.
{}From (\ref{one-exc-ene}) and (\ref{gra-rapidi})
the ground-state energy is obtained as
\begin{eqnarray}
\label{one-gro-ene}
E_N^{(0)}
&=&
\sum_{j=1}^N
\left(
k_j^{(0)}
\right)^2
=
\left(
\frac{2\pi}{L}
\right)^2
\left[
\frac{1}{3}
\lambda N^3
+
\frac{1}{2}
\lambda(\mu+\nu-\lambda)N^2
+
\frac{1}{12}
\left(
3(\mu+\nu-\lambda)^2-\lambda^2
\right)N
\right].
\end{eqnarray}
The formula (\ref{one-gro-ene}) leads to the finite-size corrections
to the ground-state energy
\begin{equation}
\label{fss-gro}
E_N^{(0)}
=
\epsilon^{(0)}L
+
2f
+
\frac{\pi v_{\mbox{{\tiny F}}}}{L}
\lambda(\Delta N_b)^2
-
\frac{\pi v_{\mbox{{\tiny F}}}}{12L}\lambda,
\end{equation}
where $\epsilon^{(0)}=4\pi^2 \lambda^2 d^3/3$, $f=
\pi^2
\lambda
(\mu+\nu-\lambda)d^2$ and
\begin{eqnarray}
\Delta N_b
=
\frac{1}{2}
\left(
1-\frac{\mu+\nu}{\lambda}
\right).
\end{eqnarray}
The quantum number $\Delta N_b$ physically represents the
phase shift due to the scattering by the impurity- and
boundary-potentials \cite{y-k-y}.
Thus the ground state which has the energy $E_N^{(0)}$ is regarded as
the phase-shifted ground state \cite{A-L94}.
If we define a hypothetical system which does not include these boundary
contributions, the corresponding ground-state
energy $\tilde{E}_N^{(0)}$ is given by
\begin{eqnarray}
\label{new-b-e}
\tilde{E}_N^{(0)}
=
E_N^{(0)} - \frac{2\pi v_{\mbox{{\tiny F}}}}{L}
\frac{\lambda}{2}(\Delta N_b)^2
=
\epsilon^{(0)}L
+
2f
-
\frac{\pi v_{\mbox{{\tiny F}}}}{12L}\lambda.
\end{eqnarray}
Comparing this with (\ref{fsscft}) one
would find a curious value of the central charge which differs from $c=1$.
The similar phenomenon was observed for the $A_{N-1}$-CS model \cite{K-Y91}.
This discrepancy may be traced back to the finite-size approximation to
the long-range potential. Then this will not affect our assumption that $c=1$.

We next calculate the finite-size corrections to the
excited states. Excited state are created as follows:
(a) we add $\Delta N$ extra particles to get the ground-state
configuration for $N+\Delta N$, and
(b) we create particle-hole excitation near the Fermi point
labeled by non-negative integers $n$.
Notice that any excitations which carry currents with
large momentum transfer are prohibited
due to the absence of translational invariance
in the $BC_N$-CS model.
Let us first create an excited state corresponding to (a).
In this case,
we can solve ABA equations (\ref{exci-rapidi}) to obtain
the pseudomomenta
\begin{equation}
\label{p-n-c}
k_j= \frac{2\pi}{L}
\left[ \lambda(N+\Delta N-j) +\frac{\mu+\nu}{2}
\right].
\end{equation}
Then the finite-size corrections to leading order in $1/L$ read
\begin{eqnarray}
\label{nun-dis}
E_{N+\Delta N}^{(0)}-E_{N}^{(0)} &\simeq&
\mu_c^{(0)} \Delta N + \frac{\pi}{L}
\left[ 4\pi\lambda(\mu+\nu-\lambda)d\Delta N
+ 4\pi\lambda^2d(\Delta N)^2\right],
\end{eqnarray}
where $\mu_c^{(0)} =\partial \epsilon^{(0)}/\partial d
=(2\pi \lambda d)^2$
is the chemical potential. This is rewritten as
\begin{equation}
\label{s-fss}
E_{N+\Delta N}^{(0)} - \tilde{E}_N^{(0)}
= \frac{2\pi v_{\mbox{{\tiny F}}}}{L}
\frac{\lambda}{2} \left(\Delta N - \Delta N_b \right)^2,
\end{equation}
where we have redefined $E_N^{(0)}$ as
$E_N^{(0)}-\mu_c^{(0)} N$.
We notice that this expression for the finite-size
spectrum is essentially the same as that for the charge sector in the
Kondo problem (see formula (49) in \cite{fky}).

The other possible low-energy excitations are provided by
particle-hole excitations (b).
The corresponding energy is obtained by adding
$2\pi v_{\mbox{{\tiny F}}}n/L$ to (\ref{s-fss}).
Hence we finally have
\begin{equation}
\label{s-fss-all}
E - \tilde{E}_N^{(0)}
=
\frac{2\pi v_{\mbox{{\tiny F}}}}{L}
\left[
\frac{\lambda}{2}
\left(\Delta N - \Delta N_b \right)^2+n \right],
\end{equation}
where $E$ stands for the energy of the excited state
specified by the set of quantum numbers $(\Delta N,\Delta N_b,n)$.

\section{Boundary Critical Exponents}

\noindent
We now wish to calculate
various critical exponents using the scaling relation (\ref{ex-corr}).
Notice that when comparing our result (\ref{s-fss-all})
with (\ref{ex-corr}) we have to replace $L$ with
$2L$ since $L$ has been defined as the periodic length of the system.

Let us first introduce an operator $\psi_b$
which corresponds to the phase-shifted ground state.
This operator can be assumed to be the boundary changing
operator\cite{A-L94} which plays a fundamental role in boundary CFT.
Then the phase-shifted ground state
is an excited state relative to $\tilde E^{(0)}_N$ defined in (\ref{new-b-e}).
The scaling dimension of $\psi_b$ is thus obtained as
\begin{equation}
x_{\psi_b}
=
\frac{L}{\pi v_{\mbox{{\tiny F}}}}
\left(
E_N^{(0)}
-
\tilde{E}_N^{(0)}
\right)
=
\frac{1}
{2\xi^2}
\left(
\Delta N_b
\right)^2,
  \label{bound-ex-gra}
\end{equation}
where we have put
$\xi=1/\sqrt{\lambda},\ \zeta=1/\sqrt{\mu+\nu}$, and hence
$\Delta N_b=(1-\xi^2/\zeta^2)/2$.

We next consider an operator
$\phi$ which induces the particle number change as well as the
particle-hole excitation
in the phase-shifted ground state. From (\ref{s-fss-all})
and (\ref{ex-corr}) we have
\begin{equation}
x_{\phi}
=
\frac{L}{\pi v_{\mbox{{\tiny F}}}}
\left(
E
-
\tilde{E}_N^{(0)}
\right)
=
\frac{1}
{2\xi^2}
\left(
\widehat{\Delta N}
\right)^2+n,
  \label{bound-ex}
\end{equation}
where
\begin{equation}
\label{modi-q-n}
\widehat{\Delta N}
=
\Delta N
-
\Delta N_b.
\end{equation}
Scaling dimensions
(\ref{bound-ex-gra}) and
(\ref{bound-ex}) take the form of conformal weights characteristic of
$c=1$ CFT. This is consistent with our starting assumption that $c=1$.
Note that it is crucial to take the fictitious ground-state energy
(\ref{new-b-e}) in order to obtain the right scaling dimensions.

The radius $R$ of compactified $c=1$ free boson is taken to be
$R=\xi$ \cite{K-Y91}. Let us concentrate on the self-dual point $R=1/\sqrt{2}$
({\it i.e.} $\lambda =2$) where the symmetry is known to be enhanced to
the level-1 $SU(2)$ Kac-Moody (KM) algebra. In the $BC_N$-CS model we have
the other continuous parameters $\mu, \, \nu$ which should also be adjusted
to achieve the $SU(2)$ point. It turns out that $\mu +\nu =0,\, 1,\, 2,\, 3$
and $4$ with $\lambda =2$ are the desired points.
Results are summarized in table \ref{su2}.
We note that several
$SU(2)$ points appeared in \cite{B-P-S} are in agreement with our result.

We have thus shown that
the low-energy critical behavior of the $BC_N$-CS model
is described by $c=1$ boundary CFT,
{\it i.e.} the universality class of a chiral Tomonaga-Luttinger liquid
where the Tomonaga-Luttinger liquid parameter $K$ is given by
$K=1/(2\xi^2)$.

\section{Discussions}

\noindent
When thinking of the application of our present results we may take into
account at least two possible physical situations depending on how the
boundary effect is switched on in the system.

{}First, suppose that we suddenly turn on the boundary effects
in the ground state just like the X-ray
absorption in metallic systems. In this case, to describe
the long-time asymptotic behavior of correlation functions, we need a set of
quantum numbers $(\Delta N,\Delta N_b,n)$ as is considered in the Kondo
problem \cite{A-L94,fky}. The boundary changing operator $\psi_b$, for
instance, is represented by $(\Delta N,\Delta N_b,n)=(0,\Delta N_b,0)$.

The other case is to compute the critical exponents of
ordinary correlation functions with boundary effects. For this we need
a set of quantum numbers $(\widehat{\Delta N},\  n)$
where $\widehat{\Delta N}$ is regarded as
the ordinary particle number change in (\ref{bound-ex})
(forgetting about $\Delta N_b$ in (\ref{modi-q-n})).
In table \ref{two-point},
we summarize boundary critical exponents for the
one-particle Green function in the above two cases
and the density-density correlation function.
It will be interesting to apply our analysis to
the Haldane-Shastry model of $BC_N$-type \cite{B-P-S} and
the dynamical (or spin) $BC_N$-CS model \cite{multi}.

In the picture corresponding to the set $(\widehat{\Delta N},\  n)$,
the $\xi$-dependent exponents appear only in the
correlation functions whose intermediate states are related
to the change of the number of particles.
Thus, the density-density correlation function
which is controlled by the excitations without the particle number change
should exhibit the asymptotic behavior given in table \ref{two-point},
that is, intermediate states contributing to this correlation function
are all particle-hole type.
This fact was confirmed in the special choice of
coupling constants \cite{Macedo} (see also \cite{y-k-y}).
This is one of the characteristic features of
chiral Tomonaga-Luttinger liquids \cite{ch-TLL}.

We finally remark that,
as the $A_{N-1}$-CS model is related to the Gaussian
random matrix theory,
the $BC_N$-CS model with appropriate coupling
constants is intimately related to
the Laguerre random matrix theory \cite{y-k-y}.
Our results on the asymptotic behavior of
correlation functions will also be useful
in the context of the Laguerre random matrix theory which finds
interesting applications in physics \cite{lrmt}.

\newpage


\noindent
{\bf Acknowledgments}
\vskip 2mm
\noindent
T.Y. was supported by
the COE (Center of Excellence) researchers program
of the Ministry of Education, Science and Culture, Japan.
N. K. was partly supported by a Grant-in-Aid from the Ministry
of Education, Science and Culture, Japan.
The work of S.-K.Y. was supported in part by Grant-in-Aid
for Scientific Research on Priority Area 231 ``Infinite Analysis'',
the Ministry of Education, Science and Culture, Japan.




\begin{table}[h]

\vskip10mm

\begin{tabular}{cll}
$\mu+\nu$ &
$x_\phi$ &
Representation\\
\tableline
2         &
$\frac{1}{4}(2\Delta N)^2 + n$  &
spin 0 rep. of
the level-1 $SU(2)$ KM algebra\\
\hline
0, 4         &
$\frac{1}{4}(2\Delta N+1)^2 + n$  &
spin 1/2 rep. of the level-1 $SU(2)$ KM algebra\\
\hline
1, 3        &
$\frac{1}{16}(4\Delta N+1)^2 + n$  &
unique rep. of the level-1 twisted $SU(2)$ KM algebra  \cite{twist} \\
\end{tabular}
\vskip5mm
\caption{$SU(2)$ points ($\lambda=2$).}
\label{su2}
\end{table}

\vskip10mm


\begin{table}[h]
\begin{tabular}{cll}
Quantum numbers  &
Two-point function &
Exponent\\
\tableline
$(\Delta N,\Delta N_b,n)
=(0,\Delta N_b,0)$              &
$\langle \psi_{b}(\tau)\psi_{b}(0) \rangle
\sim
\frac{1}{\tau^{2x_{\psi_{b}}}}$  &
$x_{\psi_b}
= \frac{1}{8\xi^2}
\left(
1
-
\frac{\xi^2}{\zeta^2}
\right)^2$\\
\tableline
$(\Delta N,\Delta N_b,n)
=(1,\Delta N_b,0)$              &
$\langle \Psi^{\dag}(\tau)\Psi(0) \rangle_{{\rm sudden}}
\sim
\frac{1}{\tau^{2x_{G}}}$  &
$x_{G}
= \frac{1}{8\xi^2}
\left(
1
+
\frac{\xi^2}{\zeta^2}
\right)^2$\\
\hline
$(\widehat{\Delta N},n)
=(1,0)$         &
$\langle \Psi^{\dag}(\tau)\Psi(0) \rangle
\sim
\frac{1}{\tau^{2x_{g}}}$ &
$x_{g}
=
\frac{1}{2\xi^2}$\\
\hline
$(\widehat{\Delta N},n)=(0,k)$        &
$\langle \rho(\tau)\rho(0) \rangle
\sim
\frac{1}{\tau^{2x}}$  &
$x=k$
 \\
\end{tabular}
\vskip5mm
\caption{Two-point correlation functions.
$\Psi$ is the particle annihilation operator,
$\rho$ is the density operator and
$\langle \cdots \rangle_{{\rm sudden}}$
stands for the expectation value
when the boundary potential is suddenly
switched on.}
\label{two-point}
\end{table}

\end{document}